\documentclass[journal]{IEEEtran}
\usepackage{amsmath,amssymb}
\usepackage[T1]{fontenc}
\usepackage{graphicx}
\usepackage{array,booktabs}
\usepackage{cite}
\usepackage{url}
\usepackage[hidelinks]{hyperref}
\providecommand{\tightlist}{\setlength{\itemsep}{0pt}\setlength{\parskip}{0pt}}
\providecommand{\ul}[1]{#1}
\renewcommand{\thetable}{\arabic{table}}
\usepackage{newunicodechar}
\newunicodechar{ξ}{\ensuremath{\xi}}
\newunicodechar{μ}{\ensuremath{\mu}}
\begin{document}
\hypersetup{pdftitle={Resource Adequacy with Renewables and Storage—Part I: An Exact Decomposition of Loss-of-Load Hours and Expected Energy Not Served},pdfauthor={Philip Q. Hanser},pdfkeywords={resource adequacy, energy storage, capacity accreditation}}
\title{Resource Adequacy with Renewables and Storage---Part I: An Exact Decomposition of Loss-of-Load Hours and Expected Energy Not Served}
\author{Philip~Q.~Hanser,~\IEEEmembership{Life~Member,~IEEE}\thanks{P.~Q. Hanser is a Principal Emeritus of The Brattle Group, Boston, MA 02108 USA (e-mail: Philip.Hanser@affiliate.brattle.com). The views expressed are the author's own and do not represent the positions of The Brattle Group or of any client. This work has been submitted to the IEEE for possible publication. Copyright may be transferred without notice, after which this version may no longer be accessible.}}
\maketitle
\begin{abstract}
Classical resource-adequacy assessment convolves a capacity-outage probability table with a load distribution. The calculation rests on two independence assumptions: among thermal forced outages, and between aggregate available capacity and demand. High renewable penetration breaks the second, because weather drives capacity and demand together, most strongly in the joint tail and differently by season. The error lies in the distribution to which the convolution is applied, not in the convolution, which remains an exact step of the correct calculation in three settings. First, netting renewable output from load reduces the problem to one dimension without loss. Second, conditional on weather factors, the convolution holds exactly inside an integral over those factors. Third, with storage, expected energy not served and loss-of-load hours for a single reservoir split exactly into two terms. The power-limited term is the classical convolution with the load axis shifted by the storage power rating. The energy-limited term is the expected cumulative energy the reservoir fails to deliver while empty, the Skorokhod regulator at its empty boundary. Added power at a fixed energy rating lowers the power-limited term but can raise the energy-limited term by nearly as much, because the reservoir empties sooner. Only added energy lowers the energy-limited term, and only where surplus energy refills the reservoir between renewable droughts. The term vanishes when the energy rating covers the largest cumulative drawdown of the droughts. A closed-form two-regime solution provides a verification benchmark. Part II demonstrates the decomposition on system data and derives its consequence for capacity accreditation.
\end{abstract}
\section*{1. Introduction}

Resource-adequacy assessment asks how much firm demand a power system can serve at a target reliability, and the calculation that has answered this question for six decades convolves a capacity-outage probability table (COPT) against a load distribution to obtain the loss-of-load and energy-not-served metrics. That calculation was designed for a fleet of independently failing thermal units serving a weather-driven but generation-independent load. Two features of the decarbonizing system violate the premises on which it rests: a large share of variable renewable energy (VRE), whose availability is driven by the same weather that drives demand, and energy-limited storage, whose contribution to adequacy depends on the path of net load rather than on its one-period distribution. The question is narrow: what part of the classical apparatus survives these two changes? The convolution survives. What does not survive is the object against which it was convolved---and, in the end, the scalar in which its answer is reported.

Two facts frame the argument. The first is the \emph{dunkelflaute}: the wide-area, multi-day lull in wind and solar output that empties reservoirs while demand runs at its heaviest. The apparatus developed here is built to measure it, because it is an energy deficit rather than a capacity deficit, counted in gigawatt-hours across days instead of gigawatts within an hour. The second is that the independent-outage tail the classical calculation quantifies has never, so far as the record shows, been the mechanism of an actual interruption. The 2003 Northeast cascade began with a failed state estimator and untrimmed vegetation. The 2011 and 2021 Electric Reliability Council of Texas events were driven by cold correlated across the footprint, which raised demand, froze thermal plant, and cut gas supply at once. Each was a common-mode or cascading failure; none was a coincidence of independent outages. The defect located here is therefore not confined to high-renewable systems. It was present in the thermal-dominated system too, concealed because the answer was never checked against an event. High penetration merely makes the neglected structure govern the ordinary case as well.

The argument turns on a distinction between an operator and its argument. The classical convolution rests on two independence assumptions of unequal fragility: independence among thermal forced outages, used to build the capacity-outage table, and independence between aggregate available capacity and demand, used to convolve that table against load. Only the second fails decisively under high VRE penetration, because the binding reliability event becomes a joint weather tail---low wind, low solar, high net demand---that is seasonally non-stationary and asymptotically tail-dependent in the way convenient parametric families are not. But that failure indicts a marginal law, not an integral. A marginal distribution of demand carries no information about the region in which demand and capacity are jointly extreme, and no care taken in the convolution recovers what the marginal discarded before the operator was reached. Replace the marginal and the operator stands.

Weather-independent thermal availability is the companion casualty, and it too has failed in most serious interruptions on record. Winterization after Winter Storm Uri loosened the coupling between cold and generator failure in Texas without breaking it, and the gas system supplying those generators remains free to curtail deliveries and declare force majeure when the temperature falls. Sections 4.2, 5, and 6 treat this at length.

Three replacements organize what follows, and each leaves a convolution exactly intact inside it. The first is the familiar one: netting renewables into load is an exact dimension reduction, because the shortfall event depends on load and renewable output only through their difference, so net load is a sufficient statistic and the dependence survives in the shape of its distribution (Section 4). The second conditions rather than pools. Renewables and load co-move because they share a few common weather drivers. Where unit availabilities are independent conditional on those drivers, the Garver construction runs exactly inside an integral over the factor law (Section 5). That law may be fitted or taken directly from the empirical weather record; the construction is the same. This is the minimal generalization of the capacity-outage table that admits correlation, and it repairs the thermal-weather coupling netting leaves open: the freeze-offs and deliverability losses that arrive in the hours net load is extreme.

The third carries the paper's principal result, and it changes the shape of the calculation rather than the shape of its argument. Once an energy-limited resource enters, the scalar net load ceases to be sufficient: the adequacy of any period depends on the state of charge, and hence on the path that produced it. The natural continuous-time representation of the state of charge is a reflected stochastic process on the interval of storable energy, regulated at its empty and full boundaries by the Skorokhod map. Within this representation the additive adequacy metrics for a single reservoir decompose exactly into two terms. The first is memoryless and power-limited: the classical convolution, with the load axis shifted by the storage power rating. The second is energy-limited, and equals the expected value of the Skorokhod regulator at the empty boundary: the cumulative energy the empty reservoir fails to deliver (Sections 7 and 8). The classical object is neither discarded nor approximated; it reappears in an exact identity, and the record of recent large interruptions indicates that it is the smaller of the two contributions.

The smaller term is also the one that can be computed exactly. The power-limited term factorizes through a marginal and requires no path information, so it can be evaluated by quadrature while simulation is confined to the boundary term, or retained as a control variate within a single simulation (Section 8.6). Part II {[}1{]} demonstrates the decomposition on 4 years of high-penetration system data and works out its consequence for capacity accreditation. The intervening sections situate these results. Section 2 states the classical construction and its two assumptions, and Section 3 locates the defect. Section 6 surveys the machinery the conditional construction draws on: extreme value theory on the joint tail, weather-consistent simulation, and rare-event techniques. Section 9 relaxes the scalar value of lost load (VOLL) into a priority-service menu, and the single boundary into a family of them; and Section 10 concludes. Appendix A supplies a closed-form empty-boundary probability for a two-regime reservoir as a verification benchmark.

\section*{2. The classical framework and its two independence assumptions}

The standard reliability calculation associated with Garver proceeds in two steps. It first convolves the two-state, or multi-state, forced-outage distributions of the individual units into a COPT. It then convolves that distribution against the distribution of demand to obtain the loss-of-load probability (LOLP) and, by summation over the horizon, the loss-of-load expectation (LOLE).\footnote{The COPT is the discrete convolution of individual unit availability distributions; LOLP is the probability that available capacity falls short of demand in a given period, and LOLE is its sum over the horizon. See {[}2{]} and {[}3{]}.} The effective load-carrying capability (ELCC) of a resource is the increment of firm demand that can be served while holding LOLE fixed when the resource is added.

Two distinct independence assumptions do work in this construction, and they are not equally fragile. The first is independence \emph{among} the thermal forced outages, used to build the COPT by convolution. The second is independence between the \emph{aggregate available capacity} and \emph{demand}, used to convolve the COPT against the load distribution. The first is defensible to a first approximation, subject to common-mode corrections. The second is the one that collapses under high VRE penetration.

\section*{3. What the unconditional convolution misses}

With a large renewable share, available capacity and demand are both weather-driven, and the binding reliability event is a joint tail event: low wind, low solar, and high (or high-shoulder) net demand occurring together. Renewable availability is spatially correlated within each technology, correlated across technologies, correlated with load---solar is anti-correlated with the evening ramp and positively correlated with daytime cooling load---and the correlation structure is non-stationary across seasons and times of day. Convolving the COPT against an independent load distribution assumes away the dependence that determines the risk.

The defect is narrow. The convolution integrates a capacity distribution against a distribution of demand; both are marginal laws, and a marginal carries no information about the region in which the two variables are jointly extreme. The dependence has not been modeled badly. It was discarded before the operator was reached. The same defect disqualifies the Gaussian copula,\footnote{Sklar's theorem states that any multivariate distribution can be written in terms of its marginals and a copula capturing the dependence structure; conversely, coupling marginals through a copula yields a valid joint distribution. See {[}4{]} and {[}5{]}.} which is asymptotically independent---its coefficient of tail dependence is zero---and therefore understates the probability of simultaneous renewable drought and high load whatever correlation matrix it is given.\footnote{Asymptotic dependence measures whether the probability of one variable being extreme, conditional on another being extreme, remains positive in the limit; Gaussian dependence forces this conditional probability to zero. See {[}6{]} for the machinery.} Nor is measured dependence weak. A principal-components decomposition of the locational capacity-factor covariance in California loads roughly four-fifths of the variance across thirteen solar sites, and more than half across forty wind sites, onto the first eigenvector.\footnote{Wolak {[}7{]} computes the contemporaneous covariance matrix of hourly locational capacity factors and reports the ratio of its largest eigenvalue to the total variance: approximately 0.80 across thirteen California solar locations, more than 0.50 across forty wind locations, and just under 0.50 across the fifty-three combined. Under cross-location independence with equal variances the ratio would equal 1/\emph{K}. The statistic measures central co-movement, not the asymptotic tail dependence that governs the shortfall region. Reproduced in {[}8{]}.} But covariance is a whole-distribution summary. A fleet can co-move strongly on average while its joint behavior in the shortfall region is set by a tail dependence the covariance does not see. That a well-estimated correlation matrix does not rescue the Gaussian copula, and a well-estimated load distribution does not rescue the convolution, are the same observation.

Two repairs follow, and they are not alternatives so much as the same move made at different levels. One can enrich the dependence model, replacing convenient families with vine (pair-copula) constructions built from \emph{tail-dependent} bivariate blocks and estimated \emph{conditionally} on season and time of day rather than pooled.\footnote{A vine, or pair-copula construction, decomposes a high-dimensional copula into a cascade of bivariate copulas, each of which may take a different family and asymmetric tail dependence. See {[}9{]}.} Or one can enrich the object convolved against, replacing the unconditional marginal with a law conditioned on the state of the weather. The second is what the remaining sections pursue, because it leaves the operator intact and inherits six decades of practice with it. Either way the correct primitive ceases to be the calendar year and becomes the stratum within which the joint law is approximately stationary; risk is estimated within homogeneous strata and aggregated. Conventional seasons are too coarse for this. The diurnal solar cycle and the seasonal temperature cycle interact, so the stratum that is approximately stationary is a month crossed with an hour of the day, or something finer still, and the count of strata rises as the data available to each falls.

Stratify finely enough and the convolution is evaluated in several hundred cells, each estimated on a handful of observations, which is neither cheaper nor more credible than driving load, wind, and solar jointly from the weather record and simulating the sequence directly. Past some degree of stratification, chronological Monte Carlo simulation with weather-dependent outage probabilities, and importance sampling for the deep tail, is the better instrument. Section 5.1 argues that it is the same construction, not a rival one: the convolution is an exact fragment of what sequential simulation computes, the fragment can be evaluated without sampling, and its size tells a planner whether the simulation was necessary. Section 8.6 makes that use explicit.

\section*{4. Recovery I: net load as a sufficient statistic}

The first recovery is the familiar one, and it is exact. Net out renewables from the load and convolve the (approximately independent) thermal COPT against the resulting net-load distribution. The maneuver is legitimate under one reading and circular under another, and the distinction is decisive.

\subsection*{4.1 An exact sufficiency argument}

Write the hourly shortfall event as available capacity less than the load, \(\{ C_{th} + R < L\}\), where \(C_{th}\) is available thermal capacity and \(R\) is available renewable output. This is algebraically identical to \(\{ C_{th} < L - R\} = \{ C_{th} < N\}\), where \(N: = L - R\) is net load. The rearrangement is exact and assumes nothing. Conditional on the realized triple, the event depends on \((L,R)\) \emph{only} through \(N\). If thermal forced outages are independent of weather, then

\[LOLP = \int F_{C_{th}}(x)\, dG(x),\]

where \(G\) is the distribution of \(N\) \emph{induced by the true joint law of} \((L,R)\). The dependence has not been discarded; it is carried in the shape of \(G\). Anti-correlation between solar and the evening ramp fattens the upper tail of \(G\), and that fat tail is the adequacy signal. That the binding event migrates off the gross-demand peak to the net-load peak is not merely formal. It is what the California shortfall of August 2020 recorded.\footnote{The California rolling blackouts of August 14--15, 2020, fell in the early-evening net-load ramp---on the order of 8,300 to 8,500 megawatts over 3 hours---rather than at the higher afternoon system-demand peak, and system demand on August 17 and 18 exceeded that of August 14, yet no curtailment occurred on those days. The binding hour sat at the net-load peak, not the gross-demand peak. See {[}8{]}.} Under weather-independent thermal availability and in the absence of storage, \(N\) is a sufficient statistic, and collapsing the bivariate object to the scalar \(N\) is lossless.

The circularity---proof by assumption---is committed in one specific way: forming \(N\) by subtracting a \emph{deterministic} or \emph{expected} solar shape, or convolving \(L\) against \(- R\) as though independent. That manufactures a thin-tailed net load and encodes the proposition that solar reliably appears in proportion to load. The correct construction subtracts at the level of joint realizations, driven by the same weather, so that the correlation survives into \(G\).

No production tool is likely to commit this error in its code, and the code is the wrong place to look for it. The error lives in the inputs. A tool that convolves a capacity-outage table against a supplied net-load distribution does its arithmetic correctly; what determines the result is how that distribution was formed. Forming it by subtracting a normalized or monthly-average renewable shape from a load forecast, instead of differencing coincident hourly realizations, is common in screening studies, in interconnection-level analyses, and in the renewable-shape assumptions behind some accreditation filings. The error is therefore made upstream of software that would compute either input faithfully, which makes it harder to find and easier to repeat. Where dependence is modeled explicitly, the pair-copula constructions of footnote 5 and the multivariate extreme-value machinery of Section 6 are the instruments for it.

Fig. 1 shows what the error costs on the data of Part II {[}1, Sec. 2.1{]}. Both constructions share a mean net load, but subtracting an expected shape manufactures a thin upper tail and erases the fat net-load tail that carries the adequacy signal. A planner setting firm capacity from the circular construction would believe the system carried a one-in-a-thousand risk while in fact carrying many times that.

\begin{figure}[!t]
\centering
\includegraphics[width=\columnwidth]{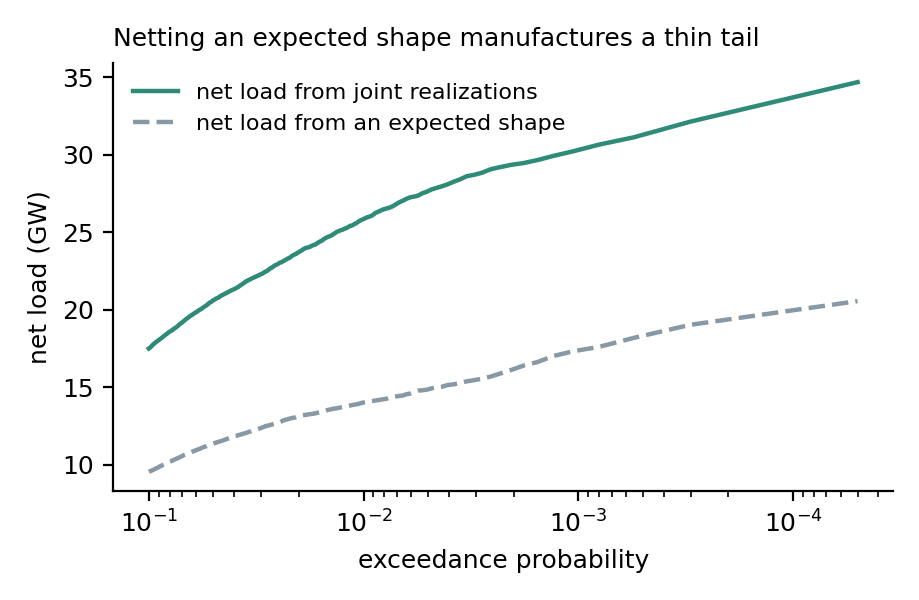}
\caption{\emph{Netting an expected renewable shape, rather than joint realizations, manufactures a thin tail and understates LOLP. Data and scenario as in {[}1, Sec. 2.1{]}.}}
\end{figure}
\subsection*{4.2 Where the reframing leaks}

Even performed correctly, netting is lossless only under conditions that are increasingly false, and those conditions locate the residual risk.

\emph{Thermal availability is not weather-independent.} The independence of \(C_{th}\) from \((L,R)\) fails badly in the cold tail---gas freeze-offs, dual-fuel and supply limits, and correlated thermal trips in the hours when net load is extreme. Netting resolves the solar--load axis and leaves this axis untouched; for a winter-peaking, high-electrification system it may be the dominant term. The hazard is not that netting is circular where it operates, but that it appears to have solved the dependence problem when it has addressed only one coordinate. Section 5 supplies the repair.

\emph{Storage breaks the sufficiency of a per-period marginal.} Once storage or any energy-limited resource enters, the adequacy of a period depends on the state of charge, and hence on the path of net load, not on its one-period distribution. The scalar \(N\) ceases to be sufficient. This is the subject of Sections 7 and 8.

\emph{Attribution dies with the scalar.} Collapsing to \(N\) is adequate for a system LOLP but useless for marginal ELCC accreditation, because it discards the multivariate object needed to decompose capacity value across resources and to expose the wind--solar diversity interaction.

\section*{5. Recovery II: conditional convolution under factor conditioning}

Section 4.2 leaves two axes open: thermal availability is not weather-independent, and storage breaks the sufficiency of a per-period marginal. The second recovery addresses the first of these, and it follows directly from the diagnosis of Section 3. If the convolution fails because the law against which it is convolved is unconditional, then condition it.

Non-stationarity is handled by conditioning rather than pooling. A regime-switching or hidden-Markov layer lets a latent weather regime govern the joint law, so that structure is never averaged across seasons. A factor model makes the same move parametrically: renewables and load co-move because they share a small set of common weather drivers---a regional temperature anomaly, a wind-resource index, a cloud-cover index---and conditional on those factors, unit availabilities are independent to a good approximation.

Let the factor state at time \emph{t} have law \emph{H}, and write \emph{F} for the conditional distribution of available thermal capacity and \emph{G} for the conditional distribution of net load, both given the realized factor value \emph{f}. The LOLP is then

\[{LOLP}_{t}\ = \ \int\ \lbrack\int\ F_{th}\left( x\ |\ f \right)\ dG_{t}\left( x\ |\ f \right)\rbrack\ dH_{t}(f),\]

Given that conditional independence, the inner integral is exactly the classical Garver convolution, evaluated on conditional laws, and the outer integral is a quadrature over the factor distribution. Nothing beyond that assumption has been approximated, and nothing has been replaced. The dependence between thermal availability and net load is carried entirely by the shared conditioning, and is therefore visible in the construction, not assumed away by it. That includes the cold-tail coupling of Section 4.2, in which the freeze-offs and deliverability losses that deepen the deficit also remove thermal capacity. This is the minimal generalization of the COPT that admits correlation while preserving the exact convolution inside the integral.

Three costs attach. The factors must be chosen so that conditional independence is credible, which is a modeling judgment, not an estimated quantity, and a factor set too coarse reinstates the defect it was introduced to repair. The conditional laws must be estimated in strata where data are thin, because the strata that bind are the rare ones. And the outer quadrature runs over a distribution whose own tail is the object of interest, so it inherits the rare-event problem that Section 6 addresses. None of these disturbs the exactness of the identity; they affect only whether it can be evaluated well.

The same costs explain a movement in practice that the identity accommodates. For the purpose of producing a number, assessment has moved away from fitting statistical models of net demand and toward using the empirical joint record---the historical weather years, or a resample of them---directly. That is not an alternative to the construction above. It is the same construction, with the factor law replaced by the empirical measure over weather years and the conditioning performed by resampling rather than by estimation. Conditioning on a fitted factor and conditioning on a realized weather year are the same move, one parametric and one not, and the nonparametric version is exact for the same reason: the joint realization carries the dependence, so nothing is discarded before the operator is reached. Two considerations favor it. Sequential simulation is required in any case once storage is material, for the reasons Sections 7 and 8 develop, so a fitted marginal buys nothing the record does not already supply. And estimation effort tends to move results little, because the metrics are dominated by a small number of weather years, so much of the fitted structure is averaged away in the quantity being reported.

What the parametric construction buys is not the number but the account of it. A resample returns a LOLE; it does not say which coordinate of the weather produced it, how far the answer would move under a climatology the record does not contain, or whether wind and demand are asymptotically dependent or merely correlated. The conditional form above is where those questions are posed, and the bivariate extreme-value work of Sanchez et al.~{[}10{]} and the modeling of wind availability conditional on temperature by Kamel et al. {[}11{]} are instances of posing them; Dent et al.~{[}12{]} situate the question within the decision the assessment is meant to support. The division of labor is clean: the empirical joint record is the right argument to the convolution when the object is a number, and the fitted conditional law is the right argument when the object is an explanation. Both are Recovery II. Only the unconditional marginal is excluded.

\subsection*{5.1 What this has to do with practice, which no longer convolves}

Resource-adequacy assessment has largely left the convolution behind. Accreditation in the large systems is computed from chronological Monte Carlo simulation over many sampled weather years, with sequential outage processes and an explicit state of charge for storage; the capacity-outage table survives as a screening device, in legacy regional studies, and inside some tools as an interim object. Against that practice, a paper reinstating the convolution would be proposing a step backward, and the objection would be decisive.

The claim here is narrower. The convolution is a term in what chronological simulation computes, exactly and not approximately, and isolating it changes what a simulation must do and what its output can mean. Three consequences follow, each a claim about sequential simulation rather than against it.

\emph{The classical object partitions the modern one.} Section 8.3 splits expected energy not served (EENS) exactly into a power-limited term and an empty-boundary term, and Part II shows the split on real data {[}1, Sec. 2.4{]}. A chronological simulation already computes their sum. Computing them separately costs nothing and tells the planner whether the risk is a shortage of megawatts or of megawatt-hours, which is the procurement question and which the aggregate metric does not answer.

\emph{The classical object is computable without sampling, so it can be removed from the variance.} The power-limited term factorizes through the one-period marginal of the deficit and needs no path information. It can be evaluated by quadrature and used as a control variate inside the sequential simulation, leaving sampling error to be paid on the boundary term alone. Section 8.6 develops this.

\emph{The classical object bounds when simulation is necessary.} Section 8.7 shows the boundary term is zero exactly when the reservoir's energy rating covers the largest cumulative energy drawdown the deficit path imposes, a statistic computed in one pass over the record. A system whose storage covers its drawdowns is memoryless, the boundary term is zero, and the classical answer is not an approximation to the right answer but the right answer. Section 8.6 states this as a screening test, and {[}1, Sec. 2.5{]} exercises it.

The same three consequences apply to ELCC, the accreditation instrument in use. Part II develops the answer, which is unfavorable but is a result, not a preference: the metric that ELCC holds fixed is a count, the term that binds is an additive functional of the path, and a count registers whether that functional is positive on a given day but not how large it is. Part II measures what that costs.

\section*{6. Adjacent machinery: extremes, simulation, and rare events}

Three further strands of the theoretical literature address the dependence and non-stationarity problems, and each supplies an ingredient the conditional construction of Section 5 requires: a model for the conditional tail, a source for the factor paths, and an estimator that samples the rare region.

Extreme value theory supplies a principled model for the tail of a distribution, fitted to the extreme observations alone; in its multivariate form it answers whether two variables are extreme together often enough to matter. Weather-consistent simulation supplies the joint realizations Section 5 requires, by driving load, wind, and solar from one weather record instead of three fitted marginals. Rare-event methods make the deep tail affordable, sampling preferentially in the region that matters and reweighting to correct the resulting bias. Coles {[}6{]} introduces the first and Resnick {[}13{]} its multivariate theory; Tillmanns et al.~{[}14{]} survey the second; Owen et al.~{[}15{]} develop the third for power systems.

\emph{Extreme value theory and asymptotic dependence.} Rather than fitting a copula over the whole distribution, one fits multivariate extreme-value models to the joint tail that causes shortfall---the upper tail of demand and the lower tail of renewable availability. Sanchez et al.~{[}10{]} fit smoothed joint distributions for demand net of renewables and for surplus/deficit, using the concept of asymptotic dependence of Section 3 to characterize the strength of co-occurrence in the extremes, and the smoothing that extreme value theory supplies mitigates the sparse-tail-data problem noted in Section 5.

The same strand absorbs the power-law formulation familiar from reliability practice. A generalized Pareto tail with positive shape parameter is regularly varying---a power law equipped with a principled threshold---so peaks-over-threshold estimation and power-law modeling are one framework rather than two, and splining an extreme-value fit into a power-law tail is the same maneuver made informally. Fitted tail-shape parameters for extreme wind speeds differ from site to site {[}16{]}--{[}18{]}, although extreme value theory places wind-speed maxima in the Gumbel domain of attraction, and the differences may reflect short records and mixed wind climates {[}17{]}, {[}19{]}. These studies concern high wind speeds, not the low-output tail that governs adequacy, so the inference here is by analogy. Under either reading, site-to-site instability is an argument not against the machinery but for applying it at the aggregation level that governs adequacy---the system residual-demand tail---rather than to site-level wind. For the joint tail, the corresponding object is multivariate regular variation. The limiting spectral measure describes how mass in the joint extremes distributes across directions, and hidden regular variation handles the asymptotically independent borderline cases a naive fit misclassifies.\footnote{A distribution is regularly varying at infinity when its survival function decays as a power law up to a slowly varying factor; a generalized Pareto tail with shape parameter ξ \textgreater{} 0 is regularly varying with tail index 1/ξ, so the heavy-tailed branch of extreme value theory and the power law coincide. Multivariate regular variation extends this to joint tails: the spectral (angular) measure characterizes the limiting distribution of directions among joint extremes, and hidden regular variation refines the theory on the subcone where full asymptotic dependence fails. See {[}13{]}.} Estimating the spectral measure confronts the same sparse-tail-data problem noted above, but it is the object that determines whether renewable droughts and demand peaks co-occur often enough to bind.

\emph{Weather-consistent sequential Monte Carlo.} Driving load, wind, and solar from one multi-decade weather record, or from a stochastic weather generator, inherits the spatial, cross-technology, resource-load, seasonal, and diurnal correlations from the data instead of imposing them parametrically. This is the nonparametric form of the conditioning of Section 5, and it is where practice has largely gone for VRE-heavy systems. The cost is that credible, internally consistent synthetic weather years with realistic inter-annual variability are difficult to generate from limited records. That cost and the sparse-tail-data problem are one difficulty seen from two sides: the practitioner's observation that adequacy metrics are dominated by a handful of weather years says, in operational terms, that the binding region holds almost no data, and that bounds what any method can deliver, fitted or empirical.

\emph{Rare-event methods and the dominant failure mode.} Because shortfall is a deep-tail event, naive simulation is expensive, and the tail is where dependence matters most. Cross-entropy and importance-sampling schemes have been developed for reliability with correlated wind and load {[}20{]}. A further strand, large-deviations analysis, recovers the \emph{most likely} joint configuration of load and renewables producing loss of load---the dominant failure mode---from the covariance structure rather than by imposition. Owen et al.~{[}15{]} develop importance sampling for the union of rare events motivated by this class of power-system problems.

The severe interruptions observed were not realizations of the coincident-but-independent forced-outage tail the capacity-outage table quantifies. The August 2003 Northeast cascade began with an energy-management-system failure and vegetation contact, then propagated. The February 2021 Texas event under Winter Storm Uri was driven by common-mode cold-induced generator trips, gas freeze-offs, and a coincident loss of gas deliverability. Independent-outage models return a nonzero LOLE, but the shortfall experienced has been dominated by correlated, common-mode, and cascading mechanisms those models omit. The implication is not that the machinery is idle but that its proper object is the correlated tail, not the independent-outage convolution.

\section*{7. Storage and path dependence}

Stated in physical terms, the calculation reduces to an energy balance once capacity is abundant and largely weather-driven. In a system with a large renewable share and material storage, installed megawatts cease to be the binding quantity. What binds is the energy balance across a sustained low-renewable episode: a \emph{dunkelflaute}, the deep renewable drought of Section 3, lasting a day or more. Four quantities govern that balance: the distribution of dunkelflaute depth and duration; the energy retrievable from fuels during the episode, net of the freeze-offs and deliverability losses an extreme-cold event itself induces; the energy carried into the episode in storage, which a forecastable event lets the operator pre-position; and the energy demanded over the episode, elevated by the same extreme weather. Peak instantaneous demand, the traditional object, recedes; cumulative energy over the correlated event replaces it. The apparatus that follows formalizes this accounting. The deficit process carries the dunkelflaute and the demand, the reservoir carries the pre-positioned energy, and fuel inventory is itself a second reservoir.\footnote{Fuel inventory---pipeline linepack, stored liquefied natural gas and oil, hydro reservoir energy---is itself an energy-limited reservoir, and the single-reservoir correspondence of Section 8 applies to it mutatis mutandis. The two reservoirs are not independent: in the cold tail the event that draws down batteries also curtails gas deliverability, so the buffers deplete together rather than diversifying. The February 2021 Texas cold event is the empirical instance, and it repeated the February 2011 event, after which federal staff recommended minimum winterization standards for gas production and processing {[}21{]}. Texas gas production fell almost 45\%, mostly through wellhead and gathering-line freeze-offs {[}22{]}. The federal review attributed 43.3\% of production declines to freezing temperatures and weather and 21.5\% to power losses at wellhead, gathering, and midstream facilities {[}23{]}. Production fell again in the Permian Basin, by about 3 billion cubic feet per day, in the storms of December 2022 and January 2024 {[}24{]}. Texas's post-2021 weatherization rules for gas facilities have been enforced lightly: of more than 7,000 facilities designated critical, 222 were cited after the 2022--23 winter, all but ten for a missed filing {[}25{]}, and smaller producers may apply for an exception for a \$150 fee {[}26{]}. Loss of power to electrically driven processing and compression was a further channel, and the industry's own accounts have emphasized it beyond its share; it is noted for completeness, not as a principal cause. What the coupling means for the apparatus is narrower than the attribution question and independent of it: the deficit process that drains the battery and the availability of the fuel buffer are driven by the same temperature, so the two reflected processes are correlated through their drivers and cannot be treated as diversifying reserves {[}8{]}. A full treatment carries a coupled, correlated pair of reflected processes; that extension is left for future work.}

\subsection*{7.1 Discrete time: the non-homogeneous Markov reward process}

With storage, the state of charge is a functional of the \emph{path} of net load, so a per-period marginal no longer suffices. In discrete time, the object is a non-homogeneous Markov reward process on the augmented state space (state of charge, weather regime), the inhomogeneity arising from diurnal and seasonal kernels that make the transition probabilities time-varying. If the dispatch rule is optimized rather than fixed, the object is a Markov decision process.

A shortfall with storage is either \emph{power-limited}---the instantaneous deficit exceeds the storage discharge power rating, independent of the reservoir level---or \emph{energy-limited}---the deficit is within the power rating, but the reservoir is empty. The power-limited part is memoryless and remains a per-instant convolution, essentially the Garver object. Only the energy-limited part is path-dependent. The two labels name which of the reservoir's two ratings binds: its discharge rate in megawatts, or its stored energy in megawatt-hours. \emph{Capacity} is reserved throughout for available generating capacity and for the megawatt quantity a capacity credit reports. Storage does not replace the classical machinery; it adds a first-passage layer on top of it.

The continuous-time formulation requires defense, since practice runs on hourly steps and a discrete chain sits closer to the data. Three reasons recommend it, and they concern representation rather than taste.

The first is that the quantity of interest is an amount of energy, and energy is the time integral of a rate. A discrete chain records the state of charge at interval ends and must attribute within-interval depletion by convention: beginning of period, end of period, or an average. The three conventions disagree by exactly the quantity this paper measures. A reservoir that empties 40 minutes into an hour has served two-thirds of that hour, and a chain stepping in hours credits it with all of the hour or with none. The continuous formulation makes within-interval depletion an object, not a rounding rule, and the regulator is what it integrates to.

The second is that the reflection is what makes the decomposition exact. In discrete time the projection onto the feasible interval is a clipping operation whose overshoot is an artifact of the step length. In continuous time it is the Skorokhod regulator, the overshoot is zero by construction, and the energy the reservoir was asked for and could not supply becomes a well-defined path functional. Section 8.5 shows the discrete recursion converging to it, which is the sense in which hourly practice approximates the continuous object rather than the reverse. The third is that the analytic results live in the continuous formulation: Appendix A, the separation of power from duration in Section 8.7, and the forward-equation characterization of the empty-boundary mass in Section 8.4 have no discrete counterparts that can be solved in closed form.

None of this asserts that storage decisions are made in continuous time. Operators schedule on intervals and the record arrives hourly. The claim is narrower: the adequacy functional is an integral, the integral is defined on paths, and the continuous representation is the one admitting a closed form and an exact decomposition. Hourly data are a quadrature of that integral. The literature drawn on is accordingly the storage and fluid-queue literature rather than the battery literature, because a battery under a reliability dispatch rule is formally a buffer: it fills on surplus, drains on deficit, and is constrained at both ends. Harrison {[}27{]} treats reflected Brownian motion as a storage model and Anick--Mitra--Sondhi {[}28{]}--{[}30{]} the regime-modulated case. What this paper adds is not a reservoir model but the identification of the empty-boundary regulator with unserved energy, and the observation that what remains once that term is removed is the classical adequacy calculation.

\subsection*{7.2 Continuous time I: the Markov-modulated fluid queue}

If generation availability lives in a finite set of regimes governed by a continuous-time Markov chain (CTMC) \(J(t)\)---an unstressed regime, a wind-drought regime, and a cold-snap regime in which thermal capacity is derated---then the state of charge is a fluid buffer whose fill and drain rate is modulated by that chain. This is the Anick--Mitra--Sondhi (AMS) model.\footnote{In the AMS model the buffer content, modulated by a finite CTMC, has a stationary distribution solving a linear system of ordinary differential equations whose solution is obtained by spectral decomposition of a generalized eigenproblem relating the drift-rate matrix and the generator. See {[}28{]}, {[}29{]}, and the survey of Kulkarni {[}30{]}. First-passage and busy-period distributions for this model are given by Palmowski {[}31{]}.} The state of charge is a \emph{regulated} (reflected) process on the interval \(\left\lbrack 0,\bar{E} \right\rbrack\) between the empty and full boundaries: it flows deterministically between regime jumps and reflects at both ends. Loss of load is the process pinned against the \emph{lower} boundary during a deficit regime---the mirror image of the overflow problem the queueing literature usually studies. The reliability metrics are boundary functionals, developed explicitly in Section 8.

The unifying framework is the piecewise-deterministic Markov process: a hybrid state (discrete regime, continuous charge), deterministic flow between jumps, and a developed hitting-probability theory for the empty state.\footnote{A piecewise-deterministic Markov process evolves by a deterministic flow punctuated by random jumps, with no diffusion term; it is the natural class for regime-driven storage. See {[}32{]}.}

\subsection*{7.3 Continuous time II: a diffusion driver and the Skorokhod problem}

Section 7.2 represented generation availability as a finite set of regimes, so that the driver of the reservoir jumped between a small number of values. In this section the driver is continuous instead: net load, or the weather anomaly behind it, is modeled as a diffusion---an Ornstein--Uhlenbeck anomaly, for instance, mean-reverting to a seasonal normal. The state of charge is then a reflected process on \(\left\lbrack 0,\bar{E} \right\rbrack\), driven by the diffusion through its rate of charge and discharge, and the reflection at the two barriers is made rigorous by the Skorokhod problem:

\[dx(t) = - \, b\left( D(t) \right)\, dt + dL_{0}(t) - dL_{\bar{E}}(t),\]

where \(L_{0}\) and \(L_{\bar{E}}\) are the boundary regulators.\footnote{The Skorokhod map takes an unconstrained path and returns the minimal-push regulated path staying within \(\left\lbrack 0,\bar{E} \right\rbrack\); the regulators \(L_{0},L_{\bar{E}}\) are non-decreasing and increase only when the process is at the respective boundary. See {[}33{]} and {[}27{]}.} \(L_{0}\), the regulator at the empty boundary, \emph{is} the cumulative unserved energy: EENS over a horizon equals \(\mathbb{E}\left\lbrack L_{0} \right\rbrack\). The noise enters through the driver and not through the state of charge itself. The state of charge is the time integral of a bounded rate, so its path is continuous and of finite variation, and the pair formed by the state of charge and its driver is a degenerate diffusion with no noise in the charge coordinate. That distinguishes the model from reflected Brownian motion, which Harrison {[}27{]} treats as a storage model and which arises as a heavy-traffic limit of the fluid queue. The distinction determines what the boundary measures. In reflected Brownian motion the regulator coincides with the local time at zero, but the process spends no time at the boundary, so the boundary term would carry energy not served and contribute nothing to loss-of-load probability. In the formulation here the process rests on the empty boundary for intervals of positive length, the regulator increases at the rate the deficit demands, and both metrics are well defined. The local time at zero, in the strict sense, vanishes for a process of finite variation {[}34{]}. The realistic object is the union of the two driver types---a background regime chain that sets drift and volatility, with a diffusion supplying within-regime variability---and the state of charge remains a reflected process of finite variation under either.

\subsection*{7.4 The reflection problem in plain terms}

The Skorokhod problem can be stated without its apparatus, and nothing in the argument below depends on the formal version.

Begin with the reservoir and ignore, for a moment, that it has a floor. Over any interval the system asks storage for some amount of energy: a positive amount when generation falls short of load, a negative amount when it exceeds load. Accumulate those requests and one obtains the level the reservoir \emph{would} have followed had it honored every request. Call it the unconstrained path. It is what a system with an infinitely deep reservoir would trace out.

The unconstrained path goes negative. It does so whenever the requests accumulated since the reservoir was last full exceed what the reservoir holds, which in a multi-day drought happens on the second or third evening. A real reservoir cannot go negative, so the physical path must depart from the unconstrained one. The Skorokhod problem specifies how it departs: by the least possible amount, and only when it must.

Picture a floor at zero and an upward push applied from underneath. The push is inactive while the path lies above the floor. It switches on only when the path is against the floor, and then supplies exactly enough to keep the path from going below it. That accumulated push is what the theory calls the \emph{regulator}, and the rule just given determines it uniquely: push only at the boundary, push only upward, push the minimum required. The Skorokhod problem states that this description determines a unique path and a unique push, and that both depend continuously on the unconstrained path one began with. Skorokhod {[}33{]} proved it. For a path in one dimension reflected at zero the construction is elementary: the accumulated push at any time is the largest deficit the unconstrained path has run up to that point, floored at zero.

This matters for adequacy because the push is not a mathematical artifact. It is the energy the system asked storage for and did not get. Every unit of upward push is a unit the reservoir was required to supply to keep the physical path non-negative and could not supply, because it was empty. A unit of demand that storage was asked to meet and did not meet is a unit of unserved energy. The regulator therefore \emph{is} the cumulative unserved energy from depletion, not a proxy for it and not proportional to it, but the same quantity written another way. The adequacy calculation then becomes a question about a boundary: the total push the floor supplied over the season, and the fraction of time it was supplying any at all. The expected regulator gives the first and yields EENS; the occupation time of the floor gives the second and yields the loss-of-load metrics.

The literature on reflected Brownian motion calls the accumulated push the \emph{local time} at the boundary, because for that process the two coincide. They do not coincide here. The state of charge has a path of finite variation, its local time at zero in the strict sense is zero, and the push is not a duration of any kind. It measures how much work the boundary had to do, accumulated over the occasions the process rested against it, and is closer to an impulse than to an elapsed time. This paper therefore calls it the \emph{regulator}, and nothing is lost by reading ``regulator at the empty boundary'' as ``cumulative energy the empty reservoir failed to deliver,'' which is what the formulas compute and what Part II reports.

Three features of the construction do work later. The push is \emph{path-dependent} in an irreducible way: whether the reservoir is empty this evening depends on the preceding several days of weather, and no distribution of this evening's net load contains that information. This is why no refinement of a per-period convolution recovers the term. The push is also \emph{additive}: 2 hours against the floor contribute the sum of what each required, and a deep hour contributes more than a shallow one. This is why a metric that counts days registers it without weighing it, which is the subject of Part II. Finally, the construction has an upper boundary, at which the push runs downward and represents curtailed surplus. That boundary matters for economics and for renewable curtailment but not for adequacy.

\subsection*{7.5 Non-stationarity as a transient boundary-value problem}

Seasonality enters rigorously here. A time-inhomogeneous generator \(Q(t)\), or time-varying drift and volatility, eliminates the stationary AMS and reflected-Brownian-motion solutions. One instead solves the \emph{transient} forward (Fokker--Planck) equation for the joint law of the state of charge and its driver, with the boundary conditions of Section 8.4, non-autonomous in \(t\), seeded by the season-entry charge distribution. This is the precise statement of why an annual stationary number is wrong: the process never reaches stationarity within the seasonal window, so one tracks the evolving law and integrates over the risk period the time-varying rate at which the empty-boundary regulator accumulates. A shorter horizon is not merely permissible but correct---the calculation is a boundary-value problem over the season.

All of the above is conditional on a dispatch policy. A reliability-greedy rule---charge on any surplus, discharge only to avert loss---makes the charge a deterministic functional of the driver path and yields the optimistic bound. For a single reservoir valued at a scalar VOLL, no other policy serves more energy on any path (Section 8.5). Economic dispatch can leave storage drawn down as it enters a scarcity event, so assessed reliability depends materially on the policy assumption; Stephen et al. {[}35{]} document the sensitivity of capacity credit to this choice.

A third policy is the realistic one, and it affects the pre-positioning term identified above. Because a dunkelflaute is a large-scale, days-ahead-forecastable weather feature, an operator anticipating one charges storage in advance of the event, drawing on dispatchable generation if necessary. In the model of Section 8 the surplus that charges storage already includes all available thermal headroom, because the deficit is measured net of available thermal capacity, so reliability-greedy dispatch keeps the reservoir as full as physical availability permits, and anticipatory dispatch cannot raise the event-entry state of charge above it. Anticipatory dispatch matters relative to economic dispatch, which draws storage down for arbitrage and may enter a scarcity event depleted. Forecast skill and lead time determine how much of the gap between economic and reliability-greedy dispatch the operator recovers, assessed reliability depends on both, and the season-entry charge distribution becomes an event-entry distribution set by the forecast.

Duration changes the character of the scheduling problem once the objective is anything other than unserved energy valued at one price. A reservoir that fills and empties on a daily cycle forgets its initial condition within a day. Long-duration energy storage (LDES), sized to ride through a multi-day dunkelflaute or to shift energy across a season, carries its state over the horizon on which risk is assessed. Under a scalar VOLL that persistence creates no opportunity cost, because a megawatt-hour discharged today serves the same megawatt-hour of load it would serve next week (Section 8.5). When the load at risk differs in value, when interruption cost rises more than in proportion to duration, or when storage is also dispatched for cost, discharging into a moderate deficit today forecloses discharging into a more costly one next week. The resulting problem carries an intertemporal opportunity cost and is formally the object hydrothermal coordination has always been: the marginal value of stored energy governs release.\footnote{In hydrothermal scheduling the marginal value of stored water---the derivative of expected future cost with respect to reservoir content---governs release: water is released when its marginal value is less than the marginal cost of the generation it displaces. The value function is obtained by backward recursion over the horizon or, where reservoirs are too numerous to discretize jointly, approximated by stochastic dual dynamic programming, which builds an outer approximation of the value function from Benders cuts sampled along forward simulation paths. See {[}36{]}. The resource-adequacy analogue replaces the displaced generation cost with the value of the load that would otherwise be curtailed. Where that value is a single number, release is never deferred (Section 8.5); where it differs across load, the optimal release rule is a set of thresholds in the state of charge (Section 9).} Evans et al.~{[}37{]} treat the allied problem of dispatching heterogeneous storage units to minimize unserved energy. How storage should be scheduled in a resource-adequacy assessment---and how much of the assessed reliability is an artifact of that choice rather than a property of the system---is an open question, and the volume of LDES now contemplated makes it a binding one. The choice is not only of a schedule but of the objective it optimizes against, since release requires a valuation of the load that would otherwise be served; Section 9 takes that up.

\section*{8. Recovery III: the boundary decomposition for a single reservoir}

The third recovery is the paper's principal result. This section works the correspondence through explicitly for one reservoir under reliability-greedy dispatch, writing LOLP, loss-of-load hours, and EENS as functionals of the empty-boundary regulator and isolating the surviving convolution term.

This section joins two notations ordinarily kept apart, the convolution notation of reliability assessment and the stochastic-process notation of reflected processes, so Table 1 collects the symbols before they are used. Two conventions hold throughout. Capital \emph{F} and \emph{G} denote distribution functions of available capacity and of net load, as in Sections 4 and 5, and never path quantities. The boundary regulators are always boundary objects even though \emph{L} alone denotes load; where ambiguity could arise the regulator is named in words.

\begin{table}[!t]
\caption{Notation used in Sections 7 and 8.}
\centering\footnotesize
\begin{tabular}{@{} >{\raggedright\arraybackslash}p{\dimexpr 0.4034\linewidth-2\tabcolsep\relax} >{\raggedright\arraybackslash}p{\dimexpr 0.3950\linewidth-2\tabcolsep\relax} >{\raggedright\arraybackslash}p{\dimexpr 0.2016\linewidth-2\tabcolsep\relax}@{}}
\toprule
Symbol & Meaning & Units \\
\midrule
\bottomrule
\(L(t)\) & load & MW \\
\(R(t)\) & available renewable output & MW \\
\(C(t),\ C_{th}\) & available thermal capacity, from the capacity-outage table & MW \\
\(N(t) = L - R\) & net load (Section 4) & MW \\
\(D(t) = N - C\) & deficit demanded of storage; negative values are surpluses (8.1) & MW \\
\(P\) & storage power rating, charge and discharge & MW \\
\(\bar{E}\) & storage energy rating & MWh \\
\(h = \bar{E}/P\) & storage duration (8.7) & h \\
\(x(t)\) & state of charge, \(x \in \left\lbrack 0,\bar{E} \right\rbrack\) & MWh \\
\(b(D)\) & power-constrained response, \(clip(D, - P,P)\) (8.1) & MW \\
\(L_{0}(t)\) & regulator at the empty boundary: cumulative unserved energy from depletion (7.4) & MWh \\
\(L_{\bar{E}}(t)\) & regulator at the full boundary: cumulative curtailed surplus & MWh \\
\(\mu(t)\) & instantaneous unserved-load rate (8.2) & MW \\
\(J(t)\) & weather and availability regime, a finite CTMC (7.2) &---\\
\(Q(t) = \left\lbrack q_{ij}(t) \right\rbrack\) & generator of \(J(t)\) & 1/h \\
\(r_{i} = - b\left( D_{i} \right)\) & drift of the state of charge in regime \(i\) & MW \\
\(p_{i}(t)\) & probability of regime \(i\) at time \(t\) &---\\
\(\pi_{i}^{0}(t)\) & empty-boundary probability atom in regime \(i\) (8.4) &---\\
\(\eta_{c},\ \eta_{d}\) & charge and discharge efficiency (8.5) &---\\
\(f,\ H\) & weather factor state and its law (Section 5) &---\\
\end{tabular}
\end{table}
\subsection*{8.1 Setup and dynamics}

Over a horizon \(\lbrack 0,T\rbrack\)---a season---let \(L(t)\) be load, \(R(t)\) available renewable output, and \(C(t)\) available thermal capacity drawn from the convolved COPT. Define net load \(N(t) = L(t) - R(t)\) and the \emph{deficit demanded of storage}

\[D(t) = N(t) - C(t) = L(t) - R(t) - C(t).\]

When \(D(t) > 0\), firm plus renewable generation falls short of load by \(D(t)\) and storage must discharge; when \(D(t) < 0\), a surplus \(\left| D(t) \right|\) is available to charge. Let the reservoir hold energy \(x(t) \in \left\lbrack 0,\bar{E} \right\rbrack\) with power rating \(P\). Define the power-constrained response

\[b(D) = clip(D, - P,P) = \max\left( - P,\ \min(P,D) \right),\]

so that on discharge the reservoir is asked for \(\min(D,P)\) and on charge it accepts \(\min\left( |D|,P \right)\). Efficiency is deferred to Section 8.5. The interior dynamics are \(\dot{x} = - \, b(D)\), and the reflected dynamics on \(\left\lbrack 0,\bar{E} \right\rbrack\) are

\[dx(t) = - \, b\left( D(t) \right)\, dt + dL_{0}(t) - dL_{\bar{E}}(t),\]

with \(L_{0},L_{\bar{E}}\) non-decreasing, \(L_{0}\) increasing only on \(\{ x = 0\}\) and \(L_{\bar{E}}\) only on \(\{ x = \bar{E}\}\).

\subsection*{8.2 Decomposition of unserved load}

Write μ(t) for the instantaneous unserved-load rate: the power demanded at time t and supplied neither by generation nor by storage, measured in megawatts, whose integral over the horizon is the energy not served. It decomposes exactly into a power-limited and an energy-limited term:

\begin{gather*}\mu(t) = \underbrace{\left( D(t) - P \right)^{+}}_{\substack{\text{power-limited}\\ \text{(memoryless)}}} + \underbrace{\frac{dL_{0}(t)}{dt}}_{\substack{\text{energy-limited}\\ \text{(path-dependent)}}},\\ \frac{dL_{0}(t)}{dt} = \min\left( D(t),P \right)\,\mathbf{1}\{ x(t) = 0\}.\end{gather*}

The energy-limited term equals the increment of the empty-boundary regulator, because at \(x = 0\) with \(D > 0\) the interior drift \(- b(D)\) pushes the reservoir below empty at rate \(\min(D,P)\), and the Skorokhod regulator \(dL_{0}\) pushes back at exactly that rate: the discharge storage was asked to supply, up to its power limit, but could not. The decomposition is verified case by case:

\begin{itemize}
\tightlist
\item
 \(x > 0,\ 0 < D \leq P\): storage covers the deficit; \(\mu = 0\).
\item
 \(x > 0,\ D > P\): storage delivers \(P\); \(\mu = D - P\) (power-limited only).
\item
 \(x = 0,\ 0 < D \leq P\): storage is empty; \(\mu = D\) (energy-limited only).
\item
 \(x = 0,\ D > P\): \(\mu = (D - P) + P = D\) (both terms; whole deficit unserved).
\end{itemize}

\subsection*{8.3 Reliability metrics as boundary functionals}

Integrating the unserved rate gives EENS:

\begin{multline*}EENS = \mathbb{E}\left\lbrack \int_{0}^{T}\mu(t)\, dt \right\rbrack\\ = \underbrace{\mathbb{E}\int_{0}^{T}\left( D(t) - P \right)^{+}\, dt}_{\substack{\text{power-limited: time-indexed}\\ \text{convolution}}} + \underbrace{\mathbb{E}\left\lbrack L_{0}(T) \right\rbrack}_{\substack{\text{energy-limited: expected}\\ \text{empty-boundary regulator}}}.\end{multline*}

The first term factorizes through the marginal law of \(D(t)\) at each instant---a time-indexed convolution of the capacity-outage table against net load, with the load axis shifted by the power rating \(P\), requiring no path information.\footnote{Interconnection enters on this side of the decomposition. A tie line is a power-limited resource shared between areas, and memoryless in the same sense. Its contribution requires the two-area law of residual demand, lagged across weather-system transit times, rather than a single-area marginal, while storage on either side retains its path dependence. The multi-area extension is left for future work; Sanchez et al.~{[}10{]} treat the two-area dependence structure.} This is the third and last place the classical convolution re-enters, and it re-enters exactly. The second term is the expected regulator of the reflected charge process at the empty boundary, and is irreducibly path-dependent.

Two qualifications recast rather than reinstate the classical object. First, the convolution re-enters only conditional on regime. The marginal of the deficit is a mixture over the weather regimes of Section 8.4, and its power-limited tail is dominated by the cold-correlated derate regime, in which thermal availability is not independent of net load: the freeze-offs and deliverability losses that deepen the deficit also remove thermal capacity. In that regime the term is not the classical independent-outage convolution at all; it is a joint object carried by the regime process, whose state set already contains the cold-snap-with-thermal-derate mode. Second, the independent-outage share of the term is empirically negligible, by the record noted in Section 6. That share is the coincident but uncorrelated thermal failures that Garver-type and Monte Carlo models compute. The operational reading is that future adequacy is carried by two terms: the energy-limited regulator term, and the correlated portion of the power-limited term routed through the regime layer. The independent-outage convolution survives formally but is small; it is retained here for completeness and for legacy systems, not because it binds.

A loss-of-load instant is any instant with \(\mu(t) > 0\), so the instantaneous LOLP is

\[LOLP(t) = Pr\{ D(t) > P\} + Pr\{ x(t) = 0,\ 0 < D(t) \leq P\},\]

and its integral over the season is the expected loss-of-load hours, \(LOLH = \int_{0}^{T}{LOLP}(t)\, dt\). The first probability is read from the marginal of \(D(t)\)---again the classical object. The second requires the \emph{joint} law of \(\left( x(t),D(t) \right)\), specifically the empty-boundary mass, which the fluid-queue or Fokker--Planck machinery supplies.

EENS and LOLH are additive functionals of the path, and both decompose exactly into the two terms. The LOLE that adequacy standards use is not additive. It counts days containing at least one loss-of-load hour, a functional of daily maxima, and a day can contain a power-limited hour and an energy-limited hour together. Its power-limited part is a lower bound on it, but it does not split into two additive terms; Part II reports it alongside the additive metrics for that reason.

\subsection*{8.4 Solving for the empty-boundary mass}

Model the regime \(J(t)\) as a finite CTMC with generator \(Q(t) = \left\lbrack q_{ij}(t) \right\rbrack\) and regime-dependent rates \(r_{i} = - \, b\left( D_{i} \right)\). Let \(F_{i}(t,x) = Pr\{ x(t) \leq x,\ J(t) = i\}\). The Markov-modulated fluid equations are

\[\frac{\partial F_{i}}{\partial t} + r_{i}\,\frac{\partial F_{i}}{\partial x} = \sum_{j}q_{ji}(t)\, F_{j}(t,x),\]

with the fluid-model boundary conditions at \(x = 0\) and \(x = \bar{E}\). Probability atoms accumulate at the boundaries in regimes whose drift points outward: the empty-boundary atom \(\pi_{i}^{0}(t) = Pr\{ x(t) = 0,\ J(t) = i\}\) builds up in deficit regimes (\(r_{i} < 0\)). A regime whose drift points away from a boundary carries no atom there; Appendix A applies both conditions to the two-regime case. The metrics then read

\begin{gather*}LOLP(t) = \sum_{i:\, D_{i} > P}p_{i}(t) + \sum_{i:\, 0 < D_{i} \leq P}\pi_{i}^{0}(t),\\ \mathbb{E}\left\lbrack L_{0}(T) \right\rbrack = \int_{0}^{T}\sum_{i:\, D_{i} > 0}\min\left( D_{i},P \right)\,\pi_{i}^{0}(t)\, dt,\end{gather*}

where \(p_{i}(t) = Pr\{ J(t) = i\}\). In the autonomous case (\(Q\) and \(r_{i}\) constant), the stationary version is solved by the AMS spectral decomposition. Under the seasonal, non-stationary generator \(Q(t)\), the transient system is integrated forward over the season from the season-entry distribution, with no spectral shortcut---the rigorous form of the claim that an annual stationary number will not do.

In the diffusion-driven formulation of Section 7.3, the corresponding object is the forward equation for the joint law of the state of charge and its driver. The equation is degenerate in the charge coordinate, because the noise enters only through the driver, and the law carries an atom on the empty boundary at positive deficits, as in the fluid model. The expected regulator is then the time integral of the power-constrained deficit, \(\min(D,P)\), weighted by the mass of that atom. It is not a probability flux, which vanishes at a reflecting boundary by construction.

\subsection*{8.5 Discrete-time reflection, efficiency, and the dispatch boundary}

The continuous regulator is the limit of a concrete discrete quantity. With step \(\Delta\), reliability-greedy dispatch gives

\[x_{k + 1} = \Pi_{\left\lbrack 0,\bar{E} \right\rbrack}\left( x_{k} - b\left( D_{k} \right)\,\Delta \right),\]

where \(\Pi\) is a projection (clipping) onto \(\left\lbrack 0,\bar{E} \right\rbrack\). The clip at zero \emph{is} the reflection, and the clipped overshoot is the discrete regulator increment---the energy storage was asked to supply but could not:

\[\Delta L_{0}^{\, k} = \left( b\left( D_{k} \right)\,\Delta - x_{k} \right)^{+}\,\mathbf{1}\{ D_{k} > 0\}.\]

The step's unserved energy is \(\left( D_{k} - P \right)^{+}\Delta + \Delta L_{0}^{\, k}\), the same power-limited-plus-energy-limited split. As \(\Delta \rightarrow 0\), the recursion converges to the reflected process and its boundary regulator.

Efficiency changes rates, not structure. With discharge efficiency \(\eta_{d}\) and charge efficiency \(\eta_{c}\), discharge drains the reservoir at the rate \(\min(D,P)/\eta_{d}\) while charging adds \(\eta_{c}\min\left( |D|,P \right)\), making the regime drift \(r_{i}\) asymmetric between charge and discharge. The reflection, the decomposition of Section 8.2, and the boundary-functional expressions of Section 8.3 are unchanged; only the drift rates and the boundary-flux scaling shift.

Dispatch policy likewise moves the boundary rather than the structure, up to a point, and that point is where long-duration storage sits. The case analysis of Section 8.2 presumes reliability-greedy dispatch: storage discharges whenever it holds energy and a deficit exists. A schedule that withholds discharge in order to preserve energy for a later, more costly hour violates that presumption, and unserved load can then occur with the reservoir neither empty nor power-limited.

For the class of policies that withhold discharge below a time- and state-dependent reserve floor, the identity survives unchanged in form: the charge process reflects at the floor instead of at zero, the energy-limited term becomes the expected regulator at the floor, and the power-limited term is untouched. Whether a given schedule belongs to that class depends on how curtailment is valued, and the simplest valuation places the floor at zero. Assume a single scalar VOLL, known and common to all load.\footnote{The VOLL is the marginal social cost of involuntary curtailment, expressed per megawatt-hour not served; it is the price at which a reliability objective converts unserved energy into a cost commensurable with generation cost, and under it the optimal release rule is to discharge whenever a deficit arises. Estimates vary across customer classes, and across the duration, timing, and notice of the interruption, by more than an order of magnitude---which is the practical form of the homogeneity objection raised in the text.} Stored energy then has no use except to avert curtailment valued at that one number, so withholding it cannot help: a megawatt-hour held back serves at most one megawatt-hour later, and the fuller reservoir spills more of any surplus that arrives in the interval. Reliability-greedy dispatch is therefore optimal path by path under a scalar VOLL, at any duration, and the floor sits at the empty boundary. That valuation is conventional, and it is the one under which Section 8.3 holds as stated, but it is an assumption and not a consequence: it presumes that curtailment is valued at a single number, that the number is known, and that the load curtailed is homogeneous in it. A positive floor requires one of those presumptions to fail, an objective that counts days instead of energy, or dispatch for cost as well as reliability.

Two difficulties attach once the floor is positive. The floor is a moving boundary, so the Skorokhod problem is posed on a time-varying domain. And the floor is an output of the scheduling problem, not an input to the reliability calculation, which couples two calculations that practice keeps separate. For policies in the threshold class neither difficulty disturbs the form of the identity; for policies outside it, a third, policy-withheld term joins the two of Section 8.3. Relaxing the scalar itself is the subject of Section 9.

\subsection*{8.6 Computational consequence: an exact fragment and a simulated remainder}

The decomposition of Section 8.3 is not only interpretive. Because the power-limited term factorizes through the marginal law of the deficit at each instant, it can be evaluated by the conditional convolution of Section 5---quadrature over the factor law, with a Garver-style discrete convolution inside---without simulating a single path. Only the boundary term requires the joint law of charge and deficit, and hence either the transient fluid or Fokker--Planck integration of Section 8.4 or a path-dependent simulation.

This suggests an estimator that sequential Monte Carlo implementations do not currently use. Standard weather-consistent simulation estimates the whole of EENS from sampled paths, and therefore pays sampling variance on both terms. The decomposition lets the power-limited term be computed exactly and the simulation confined to the boundary term. Where a single simulation is preferred, the exactly known term serves as a control variate: the path-wise power-limited energy is computable on every sampled path and has a known expectation, so subtracting it and adding back the known mean removes its contribution to the estimator's variance. The gain is bounded by the share of the variance that term carries. The argument of Section 8.3 that it is the smaller term in expectation does not settle whether it is smaller in variance: it is a deep-tail functional, and its rare, large realizations may contribute variance out of proportion to the mean. Which effect dominates is an empirical question. Part II {[}1{]} establishes the decomposition and the relative size of the two terms on a weather-consistent record; the variance decomposition, and hence the efficiency of the control variate, is not settled there and remains open. A further qualification narrows the claim. The exact evaluation integrates out thermal forced outages conditional on the weather; it does not integrate out the weather. If, as practice reports, the metrics are dominated by a few weather years, the dominant uncertainty is which years the record happens to contain. That is an uncertainty about climatology; no estimator removes it and a control variate does not touch it. The split addresses the simulation noise paid on top, which may well be the second-order term. Cutting the other way, the two terms co-move across weather years, since a year that deepens the deficit also deepens the depletion, and a control variate positively correlated with its target is more effective.

The split is also diagnostic: an assessment can report how much of its EENS and loss-of-load hours is power-limited and how much is energy-limited, which is what a system operator needs in order to know whether the answer is to procure megawatts or megawatt-hours.

That diagnostic reading returns the apparatus to the purpose for which it was built. Garver's construction was not offered as a theory of reliability but as a way to compute one, on machines that could not have carried a chronological simulation of a modern system. Its contribution was the observation that a convolution replaces an enumeration: the distribution of available capacity for a fleet of independently failing units can be built up unit by unit rather than by summing over the joint availability states. The apparatus was a device for reducing computational burden, and it succeeded so thoroughly that it came to be mistaken for the definition of the quantity it computed. The criticism of the preceding sections is aimed at that mistake and not at the device.

This suggests a use for the framework more modest, and more immediately applicable, than rebuilding an assessment around a reflected process. Adequacy assessment is expensive in a specific way: the cost sits in the sequential simulation over many weather years and outage histories, and it is paid in full whether or not the path dependence matters. Whether it matters can be estimated cheaply and in advance.

Compute the power-limited term by quadrature; this is the classical calculation with the load axis shifted by the storage power rating, and it requires no sampling. Then compute the largest cumulative energy drawdown the deficit path imposes on the reservoir, the statistic \(K\) of Section 8.7. On the weather record, with thermal capacity held at its expected availability or at a low quantile of the capacity-outage table, this takes one pass over the record and no adequacy model; across sampled outage histories it becomes a distribution. Compare the energy rating against it. A reservoir whose energy rating exceeds the drawdown carries no boundary term on that path, is memoryless for adequacy purposes, and can be assessed by the classical calculation with no loss of exactness. A reservoir whose energy rating falls short of the drawdown carries risk in the term the classical calculation cannot see, and the sequential simulation is not optional.

Three properties recommend the screen. It is one-sided in a known direction. The neglected term is non-negative, so the classical calculation is a lower bound on unserved energy: a system that fails the standard on the classical calculation fails it outright, but no classical result certifies a pass. A pass rests on the drawdown comparison, which is exact for the record in hand and inherits that record's sampling error in the drought tail. It is cheap, one ingredient a quadrature over a one-dimensional law and the other a single pass over the record. And it is informative when it fails, because the boundary term's estimated share is itself the answer to the procurement question: a system screened in for a large boundary share is being told that it needs duration rather than capacity. For a planner triaging scenarios, or a commission deciding which of a utility's cases warrant a full adequacy study, this is the practical content of the decomposition, and it requires none of the reflected-process machinery. Part II exercises it {[}1, Sec. 2.5{]}.

\subsection*{8.7 How storage power and storage duration enter}

The decomposition of Section 8.3 makes the roles of the two storage ratings explicit, and they are not the roles intuition assigns. The power rating \emph{P} enters the power-limited term, where it shifts the load axis of the convolution, and enters the boundary term only through the rate at which an empty reservoir fails to deliver. The energy rating does not enter the power-limited term at all. It enters through the boundary alone, setting how long the reservoir can absorb a deficit before the regulator switches on.

Three properties hold for any driver, not only in the two-regime case. The first is an exact criterion for the boundary term to vanish. Write \(Y(t)\) for the cumulative request \(\int_{0}^{t}b(D(u))du\) and let the reservoir start full. Until the empty-boundary regulator first acts, the state of charge equals \(\bar{E}\) less the largest rise in \(Y\) since the reservoir was last full, so the regulator stays at zero over \(\lbrack 0,T\rbrack\) if and only if

\[K = \max_{0 \leq s \leq t \leq T}\lbrack Y(t) - Y(s)\rbrack \leq \bar{E}.\]

\(K\) is the largest cumulative energy drawdown the deficit path imposes, net of the recharges within it. It is the required storage of Rippl's mass-curve method for water-supply reservoirs {[}38{]}, which Moran's probability theory of storage {[}39{]} took as its starting point, and it is computed in one pass over the record.

The second is monotonicity. Two reservoirs with the same power rating and energy ratings \({\bar{E}}_{1} < {\bar{E}}_{2}\), driven by the same deficit path and starting full, satisfy \(x_{1}(t) \leq x_{2}(t)\) at every \(t\), because the difference between the two states of charge changes only at the boundaries and stays between 0 and \({\bar{E}}_{2} - {\bar{E}}_{1}\). The larger reservoir is therefore empty only when the smaller one is, and its regulator is no larger on any path. The boundary term is non-increasing in the energy rating for every driver.

The third is a scale property. If the power rating binds in every hour, so that \(|D(t)| \geq P\) throughout, the reservoir charges and discharges at the rate \(P\) whenever it can, and the state of charge measured in hours, \(x/P\), follows a path that depends on the reservoir only through its duration \(h = \bar{E}/P\). The time spent at empty then depends on duration alone, and the energy-limited term is \(P\) times a function of duration.

The two-regime specialization of Appendix A is an instance of the scale property and gives the decay explicitly. Take the reservoir of Appendix A, alternating between a surplus regime of depth \(D_{s}\) and mean length \(1/\alpha\) and a deficit regime of depth \(D_{d}\) and mean length \(1/\beta\), and let the power rating bind in both, \(P \leq \min\left( D_{s},D_{d} \right)\). Then, with \(h = \bar{E}/P\) the storage duration and \(\kappa = \beta - \alpha\),

\[\pi_{d}^{0}(h) = p_{d}\,\frac{\kappa}{\beta\left( e^{\kappa h} - 1 \right) + \kappa},\]

which depends on the reservoir only through its duration, and not on the power rating. Appendix A.2 derives it. The governing quantity is \(\kappa h\), and since \(1/\beta\) is the mean length of a deficit episode and \(1/\alpha\) the mean length of the surplus episodes that refill the reservoir, \(\kappa h\) measures duration against drought length net of the same duration measured against recharge length. The decay it governs requires \(\kappa > 0\), which with equal charge and discharge rates means the surplus regime occupies more time than the deficit regime, so the reservoir refills on average. At \(\kappa = 0\) the boundary term falls only as \(1/(1 + \beta h)\). For \(\kappa < 0\) it does not vanish at any duration, tending to \(p_{d}(1 - \beta/\alpha)\): a system short of surplus energy needs energy, and storage of any size cannot supply it. Where \(\kappa > 0\) the term decays exponentially and never reaches exactly zero. Part II reports a term that supplies nearly all of the risk at 2 and 4 hours and vanishes identically at 16. On a finite record, exactly zero is the drawdown criterion at work: at 16 hours the energy rating exceeds the largest drawdown in every sampled outage history {[}1, Sec. 2.5{]}, although the droughts themselves last several days.

The exponential decay is a property of the Markov regime model, whose deficit episodes have exponentially distributed lengths. Weather with long memory, of the kind Hurst {[}40{]} found in Nile flow records, produces heavier-tailed drought durations, and with them a boundary term that decays more slowly in the energy rating than a Markov model fitted to the same mean drought length predicts. The drawdown criterion, computed on the record, does not depend on that model; an assessment that fits regimes does.

Two consequences follow, and both cut against the way storage is credited.

\emph{Power and energy are not substitutes.} The expected unserved-energy rate is \(p_{d}\,\left( D_{d} - P \right) + P\,\pi_{d}^{0}(h)\), so raising the power rating at fixed duration reduces it at the rate \(- \left( p_{d} - \pi_{d}^{0}(h) \right)\), which vanishes as \(\pi_{d}^{0} \rightarrow p_{d}\), that is, for a reservoir short relative to the droughts. Raising power at fixed \emph{energy rating} is worse: it lowers the power-limited term and raises the boundary term, because a reservoir permitted to discharge faster empties sooner, and the two effects nearly cancel. Beyond \(P = D_{d}\) the power rating carries no adequacy value at all, since the reservoir can already meet the whole deficit while it holds energy and every remaining shortfall is a depletion. Only stored energy retires the boundary term.

\emph{The natural sizing variable is stored energy relative to drought drawdown.} A 4-hour battery is short in a system whose droughts run for days and long in one whose binding events are evening ramps, so the same nameplate carries different adequacy value in two systems with identical load and identical thermal fleets. A capacity credit expressed as a percentage of nameplate carries none of this, which is a second sense in which the accreditation scalar is asked to summarize an object it cannot determine.

The restriction under which the separation is exact, power binding in every regime, is not the case Part II finds on the Spanish record, where the drought deficits sit mostly within the power rating {[}1, Sec. 2.3{]}. Outside it the general formula of Appendix A still holds, the power rating re-enters through the saturated rates, the separation degrades gradually rather than failing, and the drawdown \(K\), which depends on the depth of the deficit hour by hour, replaces duration as the quantity to compare.

\begin{figure*}[!t]
\centering
\includegraphics[width=\textwidth]{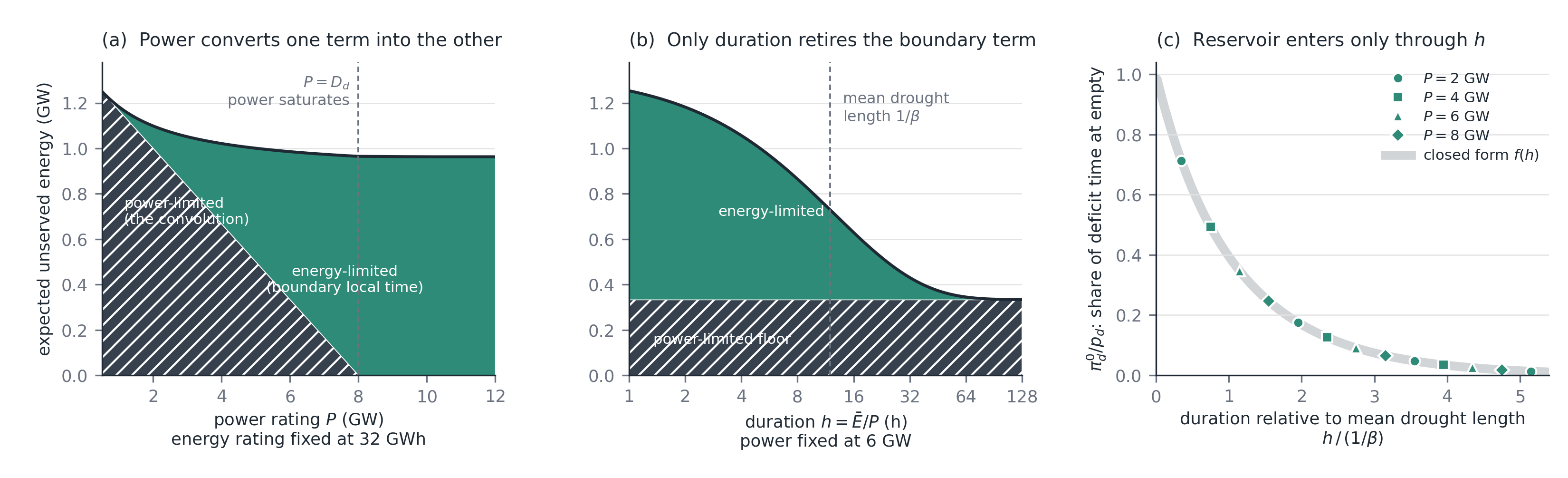}
\caption{\emph{Storage power and storage duration in the two-regime reservoir of Appendix A, with mean deficit length 12 h, mean surplus length 60 h,} \(D_{d} = 8\) \emph{GW and} \(D_{s} = 10\) \emph{GW. (a) At fixed energy rating, raising the power rating converts power-limited shortfall into energy-limited shortfall almost one for one, and has no effect beyond} \(P = D_{d}\)\emph{. (b) At fixed power rating, duration retires the boundary term, which decays as duration passes the mean drought length; the power-limited floor is irreducible in duration. (c) Within the regime in which the power rating binds, the empty-boundary probability depends on the reservoir only through its duration, and the four power ratings fall on one curve. All values are closed-form, and were verified against an event-driven simulation of the reflected process.}}
\end{figure*}
\section*{9. Voluntary curtailment: priority service and the family of boundaries}

Section 8.5 assumed a scalar VOLL in order to obtain a single reserve floor, and flagged the assumption. Relaxing it reproduces the argument of Section 3 one level down. The unconditional marginal of Section 3 fails because it collapses a joint law to a summary and discards, before the operator is reached, the dependence the shortfall region turns on. A scalar VOLL commits the same error in the same way. It collapses a distribution of valuations to a number, and it discards, before the reflection is reached, the information that determines which load is curtailed. The defect is not in the reflection. It is in the boundary.

Chao and Wilson {[}41{]} answer this by replacing administrative curtailment with economic rationing carried out by the customers themselves. The seller offers a menu of contingent forward contracts differing in service reliability; each customer selects one, and the selection fixes the order in which that customer is curtailed. The rationing is voluntary in the operative sense: no one is interrupted against a valuation an administrator imputed to them, but against a valuation they revealed by choosing and paying for it. What the menu buys is not only equity but information, since the seller learns the distribution of willingness to pay for reliability, which no behavior observed under a single price discloses. That consideration has grown rather than faded. Where the customer is a prosumer---holding generation and storage of their own, and trading off, on reasoned grounds, what to pay for grid reliability against what it costs to self-supply it---the valuation the menu elicits is a quantity the customer has already computed. The administrative scalar is an estimate of an object its owner knows and can be induced to state.

The consequence for the identity of Section 8 is not that it fails but that it multiplies. Write \emph{w(x,t)} for the marginal value of stored energy, decreasing in the state of charge; load of valuation \emph{v} is served exactly when \emph{w(x,t)} does not exceed \emph{v}, so each valuation carries its own reserve floor, the charge level at which the marginal value of stored energy rises to \emph{v}. The floors are ordered, higher-valuation load being served down to a more depleted reservoir. That ordering is not incidental: it is the monotonicity an incentive-compatible menu must satisfy.\footnote{A tariff is incentive-compatible when each customer's own selection from the menu maximizes that customer's surplus, so that none prefers the option intended for another. In the class of nonlinear tariffs to which priority service belongs, this requires the allocation to be monotone in the customer's private type---the quantity or quality assigned must be nondecreasing in willingness to pay---since otherwise a higher-valuation customer would gain by imitating a lower one. Chao and Wilson's condition that service reliability be nondecreasing in valuation {[}41, Th. 1(A){]} is that requirement specialized to reliability as the quality dimension. Wilson {[}42{]} develops the general theory and treats priority service in ch.~10.} Chao and Wilson impose it on their reliability schedule; Wilson {[}42{]} shows it to be what any nonlinear tariff requires of the allocation if self-selection is to be efficient. The decomposition of Section 8.3 then generalizes, though not as a family of regulators. The state of charge is reflected only at the lowest floor. Once it falls past a class's floor, that class is curtailed while discharge continues to serve the classes valued above it, so the charge passes through the floor instead of resting on it, and its drift changes there. Unserved energy for a class of valuation \(v\) is its share of the power-limited term, assigned to the lowest valuations first, plus the occupation integral \(\int D_{v}(t)1\{ x(t) < f_{v}\} dt\): the class's power-constrained deficit \(D_{v}\), accumulated over the time the reservoir holds less than its floor \(f_{v}\). Only the highest-valuation class, curtailed at the lowest floor, retains a regulator. The system metrics integrate these over the valuation distribution. The scalar case is the degenerate instance in which the distribution is a point mass, the floor sits at the empty boundary (Section 8.5), and the occupation integral reduces to the regulator.

The correspondence is exact at the level of their equation (12). Their service reliability is the probability that the spot price does not exceed the customer's valuation. Here the marginal value of stored energy is that price, so the reliability of a valuation class is the probability that the reservoir sits above that class's floor, and one minus it is the probability mass the forward equations of Section 8.4 place below that floor, an atom only for the highest-valuation class. Priority service prices the reservoir's shadow value; the boundary functional counts the energy it fails to release. The two are one object described from either side. Whether the pricing counterpart of the decomposition---a priority charge equal to the expectation of the marginal value of stored energy over served events, in the manner of their Theorem 1---reproduces their menu remains open.

The construction leaves two questions open. The first is the fixed point noted in Section 8.5, which the menu tightens. Chao and Wilson can treat the spot price as exogenous to any one customer, but a reservoir sets the price against which it is scheduled: the marginal value of stored energy depends on how much load the menu curtails, while the curtailment depends on that value. The second is specific to the prosumer. A customer holding storage has a state of charge, and their willingness to pay for grid reliability at any instant is a function of it---a prosumer whose battery is full values the grid less than the same prosumer late in a dunkelflaute. The valuation distribution is then not a fixed primitive but the marginal law of a second, distributed family of reflected processes, correlated with the first through the weather that depletes both. The same structure appears in the coupling of the battery and the fuel system in the cold tail (Section 7). The prosumer's reservoir is that structure at a different level of aggregation, and it too is left for future work.

The practical consequence is a change in what an assessment reports. Under a scalar boundary the metric is a single EENS, and the reliability it describes belongs to a representative customer who does not exist. Under a menu the metric is a schedule---an expected unserved energy per valuation class, each of them an occupation integral below that class's floor. That is more information, and it is the information the accreditation question of Part II requires, since what a resource is worth cannot be settled without knowing whose load it serves in the hours that bind.

\section*{10. Conclusion}

The convolution's fatal assumption is not independence among thermal outages but independence between aggregate capacity and demand, which high VRE penetration destroys through weather-driven, seasonally non-stationary joint-tail dependence. That indictment, however, falls on a marginal law rather than on an operator, and the operator survives every repair. Netting renewables into load is an exact dimension reduction when performed on the joint distribution, though it silently presumes weather-independent thermal availability and a storage-free system. Conditioning on common weather factors restores the first of those, and runs the Garver construction unchanged inside an integral over the factor law. Storage restores path dependence, whose natural continuous-time home is a Markov-modulated fluid queue or, with a diffusion driver, a reflected process governed by the Skorokhod problem. For a single reservoir, loss-of-load hours and energy not served resolve exactly into a memoryless power-limited term, the surviving convolution with its load axis shifted by the power rating, plus an energy-limited term equal to the expected regulator of the state of charge at its empty boundary.

Three limitations bound these results and mark the natural continuations. The single-reservoir correspondence treats storage and fuel inventory as one buffer; in the cold tail the battery and the fuel system deplete together rather than diversifying, so a faithful treatment carries a coupled, correlated pair of reflected processes. The decomposition is single-area; interconnection enters through the power-limited term but requires the joint, transit-lagged two-area law of residual demand, with storage on either side retaining its path dependence. And the identity is stated under reliability-greedy dispatch. It survives for schedules that withhold discharge below a reserve floor, with the reflection taken at that floor: under a scalar VOLL the floor sits at the empty boundary and reliability-greedy dispatch is optimal, and under a Chao--Wilson priority-service menu it becomes a monotone family indexed by valuation, with the energy-limited term becoming an occupation integral below each class's floor (Section 9). But the floor is a moving boundary; it is itself the output of a scheduling problem that long-duration storage turns into hydrothermal coordination; and under the menu that scheduling problem and the curtailment it prices form a fixed point. That coupling between how storage is scheduled and what reliability is then assessed to be is the most consequential of the three, because the volume of long-duration storage now contemplated makes the policy assumption, and not the system, what the metric may be measuring. None of the three disturbs the central identity. The computational claim of Section 8.6 is likewise stated and not demonstrated; Part II establishes the size of the two terms but not their relative contribution to sampling variance, so whether confining simulation to the boundary term repays the quadrature it requires remains open.

Two of the results are usable without adopting any of the apparatus. The first is the sizing criterion. The term the classical calculation cannot see is zero exactly when the reservoir's energy rating covers the largest cumulative energy drawdown in the record, so a planner can tell from one pass over a weather record and a nameplate whether the omission matters, and Part II confirms the criterion on real data. The second is the screen that follows from it. The power-limited term is a quadrature and costs nothing; compared against that criterion it says whether a sequential simulation is required. Because the term it omits is non-negative, a failing classical result is conclusive and a passing one is not; the pass rests on the drawdown comparison. Garver's construction was offered as a way of making an adequacy calculation affordable, and the decomposition returns it to that office---not as the calculation, but as the part of the calculation available in closed form, and as the instrument that says how much is left over.

For assessment practice the implication inverts the usual reading. The convolution that has served resource adequacy since Garver is not a casualty of the energy transition; it is the one part of the apparatus that has needed no repair. The casualty is the scalar it reports into. A capacity credit is defined to hold fixed a count of days, and a count of days registers whether the boundary term is positive but not how large it is; so accreditation misstates the binding term by construction, by a factor the decomposition predicts from the boundary term's share. What must be rebuilt is what sits on either side of it---the law it integrates against, which must be conditional and jointly realized, and the path-dependent term it omits, which in a high-renewable, high-storage system is the term that binds. The adequacy metrics, and the capacity-accreditation frameworks built on them, must accordingly be computed on the reflected process; but the classical calculation sits inside that computation, exact and unchanged, as the fragment for which the profession already has six decades of data.

\section*{Acknowledgment}

The author used a large language model (Anthropic Claude) to assist with reformatting the manuscript to IEEE style, with language editing, and with drafting connective and framing text. All technical results, derivations, and conclusions are the author's own and were verified by the author.

\section*{Declaration of competing interest}

The author is a Principal Emeritus of The Brattle Group, an economic consultancy whose clients include regulated utilities, regulators, and other participants in electricity markets. The views expressed are the author's own and do not represent the positions of The Brattle Group or of any client. The author received no specific funding for this work and declares no further competing interests.

\section*{Data availability}

This paper is theoretical; no new data were generated. The illustrative eigenvalue statistics cited in Section 3 are drawn from the published sources referenced there, and the closed-form benchmark of Appendix A is fully specified in the text. The numerical demonstration and its data and code accompany Part II.

\section*{Appendix A. Closed-form empty-boundary probability for a two-regime reservoir}

This appendix specializes Section 8.4 to two regimes and solves the stationary Anick--Mitra--Sondhi system in closed form, yielding an explicit empty-boundary (loss-of-load) probability against which a numerical implementation of the general system can be checked. The two-regime specification is a verification benchmark, not a modeling claim: nothing in Sections 7 and 8 requires that the system present two clean regimes, and the factor-conditioned and diffusion-driven formulations dispense with discrete regimes entirely.

\textbf{Setup.} Let the regime process \(J(t) \in \{ s,d\}\) alternate between surplus (\(s\)) and deficit (\(d\)), with \(s \rightarrow d\) at rate \(\alpha\) and \(d \rightarrow s\) at rate \(\beta\), so that

\[Q = \begin{pmatrix}
 - \alpha & \alpha \\
\beta & - \beta
\end{pmatrix},\quad\quad p_{s} = \frac{\beta}{\alpha + \beta},\quad p_{d} = \frac{\alpha}{\alpha + \beta}.\]

In surplus, the reservoir charges at the power-constrained rate \(c = min\left( \left| D_{s} \right|,P \right) > 0\); in deficit it discharges at a rate \(e = min\left( D_{d},P \right) > 0\). Assume \(D_{d} \leq P\), so the deficit is energy-limited and the sole loss-of-load channel is depletion. The drift matrix is \(R = diag\left( r_{s},r_{d} \right) = diag(c, - e)\).

\textbf{Spectral solution.} Writing \(F(x) = \left( F_{s}(x),F_{d}(x) \right)\) with \(F_{j}(x) = Pr\{ X \leq x,\, J = j\}\), the stationary system is \(F\prime(x)R = F(x)Q\). Seeking modal solutions \(F(x) = \phi\, e^{zx}\) gives the generalized eigenproblem \(\phi(Q - zR) = 0\), whose characteristic equation is

\[\det(Q - zR) = z\left\lbrack (\beta c - \alpha e) - ce\, z \right\rbrack = 0,\]

with roots

\[z_{0} = 0,\quad\quad z_{1} = \frac{\beta c - \alpha e}{ce} = \frac{\beta}{e} - \frac{\alpha}{c}.\]

The associated left eigenvectors are \(\phi^{(0)} \propto (\beta,\alpha)\)---the stationary regime vector---and \(\phi^{(1)} \propto (e,c)\). The general solution is therefore

\[F(x) = a_{0}\,(\beta,\alpha) + a_{1}\,(e,c)\, e^{z_{1}x},\quad\quad x \in \left\lbrack 0,\bar{E} \right\rbrack.\]

\textbf{Boundary conditions.} A phase carries a probability atom at a boundary only if its drift points into that boundary. At \(x = 0\) the surplus phase has positive drift and is pushed away, so it carries no atom: \(F_{s}(0) = 0\). At \(x = \bar{E}\) the deficit phase has negative drift and is pushed away, so it reaches its full probability continuously: \(F_{d}\left( \bar{E} \right) = p_{d}\). These two conditions determine the constants,

\[a_{0}\beta + a_{1}e = 0,\quad\quad a_{0}\alpha + a_{1}c\, e^{z_{1}\bar{E}} = p_{d},\]

giving \(a_{0} = - a_{1}e/\beta\) and

\[a_{1} = \frac{p_{d}\,\beta}{\beta c\, e^{z_{1}\bar{E}} - \alpha e}.\]

The empty-boundary atom in the deficit phase---the quantity of interest---is \(\pi_{d}^{0} = F_{d}(0) = a_{0}\alpha + a_{1}c = a_{1}(\beta c - \alpha e)/\beta = a_{1}z_{1}ce/\beta\). Substituting and simplifying with \(\alpha e = c\left( \beta - z_{1}e \right)\) yields the closed form

\[\pi_{d}^{0} = Pr\{ X = 0,\, J = d\} = \frac{p_{d}\, z_{1}\, e}{\beta\left( e^{z_{1}\bar{E}} - 1 \right) + z_{1}\, e}.\]

For this stylized model \(\pi_{d}^{0}\) is the instantaneous LOLP, and the expected energy-not-served rate is \(e\,\pi_{d}^{0}\), so that \(EENS = e\,\pi_{d}^{0}\, T\) in steady state over a horizon of length \(T\).

\textbf{Checks.} Two limits confirm the expression. As \(\bar{E} \rightarrow 0\) (no storage), \(e^{z_{1}\bar{E}} \rightarrow 1\) and \(\pi_{d}^{0} \rightarrow p_{d}\): with no reservoir, the system is empty whenever it is in deficit, so the LOLP collapses to the deficit-regime probability. As \(\bar{E} \rightarrow \infty\) with net-positive mean drift---\(z_{1} > 0\), equivalently \(p_{s}c > p_{d}e\)---one has \(e^{z_{1}\bar{E}} \rightarrow \infty\) and \(\pi_{d}^{0} \rightarrow 0\): a large reservoir that fills on average is almost never empty. If instead \(z_{1} < 0\), then \(e^{z_{1}\bar{E}} \rightarrow 0\) and \(\pi_{d}^{0} \rightarrow p_{d}z_{1}e/(z_{1}e - \beta) > 0\): a reservoir that does not fill on average is empty a positive fraction of the time however large it is. In the balanced case \(z_{1} \rightarrow 0\) (zero mean drift), the limit is finite, \(\pi_{d}^{0} = p_{d}\, e/\left( e + \beta\bar{E} \right)\), which again returns \(p_{d}\) as \(\bar{E} \rightarrow 0\). A solver for the general multi-regime system of Section 8.4 should reproduce this \(\pi_{d}^{0}\) to within discretization error on the two-regime instance before being trusted on the full fleet.

\textbf{A.2 Dependence on storage power and storage duration.} Section 8.7 states that the power rating and the energy rating act on different terms. The two-regime model makes this exact. Retain the notation above, write \(c = \min\left( D_{s},P \right)\) and \(e = \min\left( D_{d},P \right)\) for the power-constrained charge and discharge rates, and relax \(D_{d} \leq P\) so the deficit may exceed the power rating. The unserved-energy rate is then the sum of a power-limited term \(p_{d}\left( D_{d} - P \right)^{+}\) and an energy-limited term \(e\,\pi_{d}^{0}\). The instantaneous LOLP is \(p_{d}\) when \(D_{d} > P\), every deficit instant being a shortfall instant whatever the reservoir holds, and \(\pi_{d}^{0}\) otherwise.

Now restrict to the case in which the power rating binds on both sides, \(P \leq \min\left( D_{s},D_{d} \right)\), so \(c = e = P\). The eigenvalue becomes \(z_{1} = \beta/e - \alpha/c = \kappa/P\) with \(\kappa: = \beta - \alpha\), so \(z_{1}e = \kappa\) and, writing \(h = \bar{E}/P\), \(z_{1}\bar{E} = \kappa h\). Substituting into the closed form above,

\[\pi_{d}^{0} = \frac{p_{d}\, z_{1}\, e}{\beta\left( e^{z_{1}\bar{E}} - 1 \right) + z_{1}e} = \frac{p_{d}\,\kappa}{\beta\left( e^{\kappa h} - 1 \right) + \kappa},\]

in which the power rating cancels identically. The empty-boundary probability depends on the reservoir only through its duration, and since \(1/\beta\) is the mean length of a deficit episode, the governing group is duration measured against the mean length of the renewable droughts. In the balanced case \(\kappa \rightarrow 0\) the limit is \(\pi_{d}^{0} = p_{d}/(1 + \beta h)\), with the same property. For \(\kappa < 0\) the probability tends to \(p_{d}(1 - \beta/\alpha)\) as duration grows, and no duration retires the term.

Two comparative statics follow. Differentiating the unserved-energy rate in \(P\) at fixed duration gives \(- \left( p_{d} - \pi_{d}^{0}(h) \right) \leq 0\), which vanishes as \(\pi_{d}^{0} \rightarrow p_{d}\): adding power to an energy-starved system buys almost nothing. Differentiating in \(h\) at fixed power gives a strictly negative derivative decaying exponentially in \(\kappa h\), so duration retires the boundary term and does so faster once duration exceeds drought length. At fixed energy rating, raising \(P\) lowers the power-limited term and raises the boundary term, since \(h = \bar{E}/P\) falls; Fig. 2(a) shows these nearly canceling, and that beyond \(P = D_{d}\) the power rating carries no adequacy value.

These expressions were verified against an event-driven simulation of the reflected process over 4 million hours, agreeing to within Monte Carlo error across the parameter grid of Fig. 2.

\end{document}